\documentstyle[12pt,epsf,psfig,amsmath,axodraw]{article}

\def\Year{\expandafter\eatPrefix\the\year}
\newcount\hours \newcount\minutes
\def\monthname{\ifcase\month\or
January\or February\or March\or April\or May\or June\or July\or
August\or September\or October\or November\or December\fi}
\def\shortmonthname{\ifcase\month\orx
Jan\or Feb\or Mar\or Apr\or May\or Jun\or Jul\or
Aug\or Sep\or Oct\or Nov\or Dec\fi}

\def\TimeStamp{\hours\the\time\divide\hours by60%
\minutes -\the\time\divide\minutes by60\multiply\minutes by60%
\advance\minutes by\the\time%
${\rm \shortmonthname}\cdot   \if\day<10{}0\fi\the\day\cdot   \the\year
\qquad\the\hours:\if\minutes<10{}0\fi\the\minutes$}

\newskip\humongous \humongous=0pt plus 1000pt minus 100pt
\def\caja{\mathsurround=0pt}
\def\eqalign#1{\,\vcenter{\openup1\jot \caja
       \ialign{\strut \hfil$\displaystyle{##}$&$
        \displaystyle{{}##}$\hfil\crcr#1\crcr}}\,}
\newif\ifdtup

\newcounter{eqnumber}[section]
\renewcommand{\theeqnumber}{\thesection.\arabic{eqnumber}}
\def\equn{\refstepcounter{eqnumber}
\eqno({\rm \theeqnumber})
}

\def\npb#1#2#3{{\rm Nucl. Phys. B}{\bf \ #1}, #3 (#2)}

\def\cqg#1#2#3{{\rm Class. and Quant.\ Grav.} {\bf  #1}, #3 (#2)}

\def\hepth#1{[hep-th/#1]}
\def\hepph#1{[hep-ph/#1]}

\newbox\charbox
\newbox\slabox
\def\s#1{{      
        \setbox\charbox=\hbox{$#1$}
        \setbox\slabox=\hbox{$/$}
        \dimen\charbox=\ht\slabox
        \advance\dimen\charbox by -\dp\slabox
        \advance\dimen\charbox by -\ht\charbox
        \advance\dimen\charbox by \dp\charbox
        \divide\dimen\charbox by 2
        \raise-\dimen\charbox\hbox to \wd\charbox{\hss/\hss}
        \llap{$#1$}
}}

\def\spa#1.#2{\left\langle#1\,#2\right\rangle}
\def\spb#1.#2{\left[#1\,#2\right]}
\def\lor#1.#2{\left(#1\,#2\right)}

\catcode`@=11  

\def\Slash#1{\hskip 0.05 cm \slash\hskip -0.22 cm #1}

\def\eps{\epsilon}

\def\la{\langle}
\def\ra{\rangle}

\def\lsl{\not{\hbox{\kern-2.3pt $\ell$}}}
\def\ksl{\not{\hbox{\kern-2.3pt $k$}}}

\def\rg{r_{\Gamma}}

\def\spa#1.#2{\left\langle#1\,#2\right\rangle}
\def\spb#1.#2{\left[#1\,#2\right]}
\def\lor#1.#2{\left(#1\,#2\right)}

\def\sand#1.#2.#3{%
  \left\langle\smash{#1}{\vphantom1}\right|{#2}%
  \left|\smash{#3}{\vphantom1}\right\rangle}
\def\sandp#1.#2.#3{%
  \left\langle\smash{#1}{\vphantom1}^{-}\right|{#2}%
  \left|\smash{#3}{\vphantom1}^{+}\right\rangle}
\def\sandpp#1.#2.#3{%
  \left\langle\smash{#1}{\vphantom1}^{+}\right|{#2}%
  \left|\smash{#3}{\vphantom1}^{+}\right\rangle}
\def\sandmm#1.#2.#3{%
  \left\langle\smash{#1}{\vphantom1}^{-}\right|{#2}%
  \left|\smash{#3}{\vphantom1}^{-}\right\rangle}
\def\sandpm#1.#2.#3{%
  \left\langle\smash{#1}{\vphantom1}^{+}\right|{#2}%
  \left|\smash{#3}{\vphantom1}^{-}\right\rangle}
\def\sandmp#1.#2.#3{%
  \left\langle\smash{#1}{\vphantom1}^{-}\right|{#2}%
  \left|\smash{#3}{\vphantom1}^{+}\right\rangle}

\def\Aloop{A^{\rm 1-loop}}

\def\Mloop{M^{\rm 1-loop}}
\def\Mtree{M^{\rm tree}}

\def\Atree{A^{\rm tree}}
\def\Aloop{A^{\rm 1-loop}}

\def\BR#1#2{\la#1|{K_{abc}}|#2\ra}
\def\BRTTT#1#2{\la#1^+|\Slash{K}_{abc}|#2^+\ra}

\def\tree{{\rm tree}}

\def\NeqEight{{\cal N} = 8}
\def\NeqFour{{\cal N} = 4}

\begin{document}

\begin{titlepage}

\begin{flushright}
\today
\\
hep-th/0503102
\\
SWAT-05-426 \\

\end{flushright}

\vskip 2.cm

\begin{center}
\begin{Large}
{\bf 
Six-Point One-Loop $\NeqEight$ Supergravity   NMHV Amplitudes and
their IR behaviour}

\vskip 2.cm

\end{Large}

\vskip 2.cm
{\large
N.~E.~J.~Bjerrum-Bohr${}$,
David~C.~Dunbar${}$ and
Harald~Ita${}$
} 

\vskip 0.5cm

\vskip 0.5cm

{\it  Department of Physics \\
University of Wales Swansea \\
 Swansea, SA2 8PP, UK }

\vskip .3cm

\begin{abstract}
We present compact formulas for the box coefficients of the six-graviton
NMHV one-loop amplitudes in $\NeqEight$ supergravity.
We explicitly
demonstrate that the corresponding box integral functions, with these
coefficients, have the complete IR singularities expected of the
one-loop amplitude.
This is strong evidence for the conjecture that $\NeqEight$
one-loop amplitudes may be expressed in terms of
scalar box integral functions.
This structure, although unexpected from a
power counting viewpoint, is analogous to the structure of
$\NeqFour$ super-Yang-Mills amplitudes.
The box-coefficients match the tree amplitude terms arising from
recursion relations.

\end{abstract}

\end{center}

\vfill

\end{titlepage}

\section{Introduction}

Maximal $\NeqEight$ supergravity~\cite{ExtendedSugra} is a remarkable theory,
rich in symmetries, however, the knowledge of its perturbative expansion is
relatively poor compared to $\NeqFour$ super-Yang-Mills. In determining
the one-loop amplitudes of $\NeqFour$ SYM, a key observation is that the amplitudes
may be expressed in terms of scalar box-integral functions with rational
coefficients~\cite{BDDKa,BDDKb},
$$
 \Aloop= \sum_{a} \hat c_a I_a\,.
\equn\label{OnlyBoxesEQ}
$$
Considerable progress has recently been made in determining the coefficients,
$\hat c_a$, using a variety of methods including those based on
unitarity~\cite{BDDKa,BDDKb,BeDeDiKo,BDKn} and those inspired by the weak-weak
duality~\cite{Witten:2003nn} between $\NeqFour$ Yang-Mills and a twistor string
theory~\cite{Cachazo:2004dr,Britto:2004nj,Bidder:2004tx,BrittoUnitarity}.

String theory, inspires a relation between the Yang-Mills amplitudes and
those of gravity at tree level,
$$
\hbox{gravity} \sim \hbox{(gauge theory)} \times \hbox{(gauge theory)} \,,
\equn\label{GravityYMRelation}
$$
which arises from the heuristic relation
$$
\hbox{closed string} \sim \hbox{(left-moving open string)} \times
                          \hbox{(right-moving open string)} \,.
\equn\label{StringRelation}
$$
This has a concrete realisation in the Kawai, Lewellen and Tye (KLT)
relations~\cite{KLT} which express gravity tree amplitudes in
terms of quadratic products of Yang-Mills tree amplitudes.
Even in low energy effective field theories for
 gravity~\cite{Donoghue:1994dn} the KLT-relations
can be seen to link effective operators~\cite{EffKLT},
and KLT-relations also hold regardless of massless
matter content~\cite{BDWGravity}.

At one-loop level,  string theory would suggest such a relation {\it
within} the loop momentum integrals.
Such relations would not be
expected to persist in the amplitude after integration have been performed.
The first definite calculation of a one-loop $\NeqFour$ amplitude
was performed by Green, Schwarz and Brink~\cite{GSB}, who
obtained the four point one-loop amplitude
$$
\Aloop(1,2,3,4)
=st \times \Atree(1,2,3,4) \times I_4(s,t)\,.
\equn
$$ Here $I_4(s,t)$ denotes the scalar box integral with attached legs in
the order $1234$
and $s$, $t$ and $u$ and the usual Mandelstam variables.
The $\Aloop$ and $\Atree$ are the color-ordered
partial amplitudes.  Similarly they computed
 the one-loop $\NeqEight$ amplitude to be\,\footnote{In this, and in the following, we suppress a factor of $(\kappa/2)^{(n-2)}$
in the $n$-point tree amplitude and a factor of $(\kappa/2)^{n}$
in the $n$-point one-loop amplitude.}
$$
\Mloop(1,2,3,4)
=stu\times \Mtree(1,2,3,4)\times \Bigl( I_4(s,t)+I_4(s,u)+I_4(t,u)
\Bigr)\,,
\equn
$$
so that, like the $\NeqFour$ Yang-Mills amplitude, the $\NeqEight$
amplitude can be expressed in terms of scalar box-functions.

From a power counting analysis, the similarities
of the four-point amplitudes, are not expected to extend
to higher point functions.
In evaluating loop amplitudes in a gauge theory
one must perform integrals over the loop
momenta, $\ell^\mu$, with polynomial numerator  $P(\ell^\mu)$.
In a Yang-Mills theory,
the
loop momentum polynomial is generically of degree $n$ in a $n$-point
loop. The $\NeqFour$ one-loop amplitudes have a considerable
simplification where the loop momentum integral is of degree
$n-4$~\cite{BDDKa}. 
Consequently,  the amplitudes can
be expressed as a sum of scalar box integrals with rational
coefficients,
as follows from a Passarino-Veltman
reduction~\cite{PassVelt}.\,\footnote{ The Passarino-Veltman reduction
expresses an $n$-point integral with loop momentum polynomial of degree $m$ as
a sum of $n-1$-point integrals with loop momentum
polynomials of degree $m-1$.
For $\NeqFour$, since the loop momentum polynomial for the $n$-point
is only  of degree $n-4$ , repeated
Passarino-Veltman reductions express the amplitude as a sum of scalar boxes.}
The equivalent power counting arguments for
supergravity~\cite{GravityStringBased} give the loop momentum
polynomial of an $n$-point amplitude as having degree
$$
2(n-4)
\equn
$$ consistent with the heuristic relation
eq.~(\ref{GravityYMRelation}).
Performing a Passarino-Veltman
reduction for $n>4$
would lead one to express the amplitude as a sum of tensor box integrals
with non-trivial integrands of degree $n-4$.
These tensor integrals would be expected to reduce to scalar
boxes {\it and}  triangle, bubble and rational functions.

Despite this power counting argument, there is evidence that the
one-loop amplitudes of $\NeqEight$ can be expressed simply as a sum
over scalar box integrals analogous to the $\NeqFour$
case~(\ref{OnlyBoxesEQ}). Triangle or bubble functions do not appear
in any computation. Neither do factorisation properties of the
physical amplitudes demand the presence of these functions.  In
ref.~\cite{Bern:1998sv} the five and six point amplitudes were
computed and an all-$n$ form of the supergravity ``Maximally Helicity
Violating'' (MHV) amplitudes was presented. The simplification is peculiar to
$\NeqEight$ and explicitly does not apply for ${\cal N} < 8$ supergravities
~\cite{Grisaru:1979re,GravityStringBased,Dunbar:1999nj}.
 The similarity between 
$\NeqFour$ and $\NeqEight$  amplitudes is also apparent in the two-loop four point amplitude~\cite{BDDPR,Howe:2002ui}.  These forms for the
amplitude consist only of box functions.  Recently, it was conjectured
in~\cite{BeBbDu} from factorisation arguments that all one-loop
$\NeqEight$ supergravity amplitudes can be expressed as only box
functions and the coefficients of the boxes were given for the
six-point NMHV one-loop amplitude.

In this letter we will explore this conjecture further. First, we
present the box coefficients in an equivalent but more compact
form. Next, we will show that boxes contain the entire
IR singularities expected in the one-loop amplitude.
Other integral functions such as scalar triangle functions
generically contain IR singularities.  If such functions were present,
IR singularities would have to cancel between these functions alone.
Since the five-point amplitude only contains box integral functions
these other functions would
also have to conspire to not contribute to any of the 
soft or factorisation limits or generate UV singularities.
This provides
strong evidence for the absence of other integral 
functions for general $n$-point amplitudes.

In Yang-Mills theory, an analysis of the IR singularities led to
compact forms of the tree amplitudes. We will also use the
box-coefficients to determine forms of the amplitude and compare these
to the forms derived recently using recursive 
techniques~\cite{Bedford:2005yy,Cachazo:2005ca}.

\section{Box Coefficients of $\Mloop(1^-,2^-,3^-,4^+,5^+,6^+)$}

Quadruple cuts were used in \cite{BrittoUnitarity} to compute the
box coefficients of $\NeqFour$ gauge theory algebraically from the tree amplitudes.
As shown in~\cite{BeBbDu}, this technique, together with the KLT
relations allows the computation of box-coefficients in $\NeqEight$ supergravity
amplitudes. In particular box
coefficients were given for the ``next-to-MHV'' (NMHV) six-point
amplitude $M(1^-2^-3^-4^+5^+6^+)$, the simplest non-trivial
non-MHV amplitude.

We will use these methods and compute the explicit form of the box
coefficients. We will present the box coefficients in a very compact
form, which allows to point out similarities of $\NeqEight$ supergravity and
$\NeqFour$ super-Yang-Mills.

The six-point NMHV amplitude contains two types of box integral functions:
the one-mass box and the ``two mass-easy'' box,

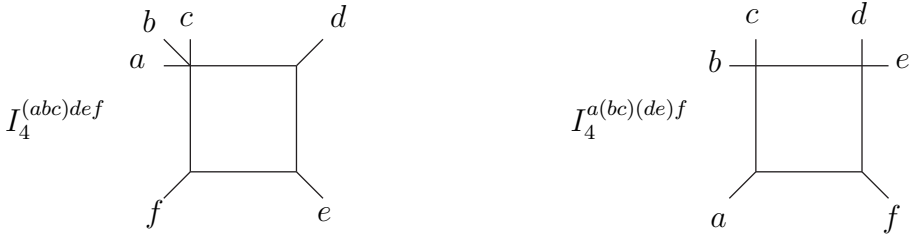
\begin{center}
\begin{picture}(150,110)(-10,0)

\Line(30,30)(30,70)
\Line(70,30)(70,70)
\Line(30,30)(70,30)
\Line(70,70)(30,70)

\Line(30,30)(20,20)
\Line(70,30)(80,20)

\Line(30,70)(20,70)
\Line(30,70)(20,80)
\Line(30,70)(30,80)

\Line(70,70)(80,80)

\Text(13,14)[l]{$f$}
\Text(78,14)[l]{$e$}

\Text(7,72)[l]{$a$}
\Text(12,87)[l]{$b$}
\Text(26,88)[l]{$c$}

\Text(83,89)[l]{$d$}

\Text(-40,50)[l]{$I_4^{(abc)def}$}

\SetWidth{1.5}

\end{picture}
\begin{picture}(150,100)(-70,0)

\Line(30,30)(30,70)
\Line(70,30)(70,70)
\Line(30,30)(70,30)
\Line(70,70)(30,70)

\Line(30,30)(20,20)
\Line(70,30)(80,20)
\Line(30,70)(20,70)
\Line(30,70)(30,80)

\Line(70,70)(70,80)
\Line(70,70)(80,70)

\Text(13,12)[l]{$a$}
\Text(78,12)[l]{$f$}
\Text(12,72)[l]{$b$}
\Text(26,89)[l]{$c$}
\Text(66,90)[l]{$d$}
\Text(83,72)[l]{$e$}

\Text(-40,50)[l]{$I_4^{a(bc)(de)f}$}
\end{picture}
\end{center}
and may be expressed in terms of these as
$$
\Mloop(1^-2^-3^-4^+5^+6^+)|_{boxes}=
c_{\Gamma} 
\biggl( 
\sum_{(abcdef)\in P_6''}  \hat c^{\,(abc)def} I_4^{(abc)def}
+
\sum_{(abcdef)\in P_6'}  \hat c^{\,a(bc)(de)f} I_4^{a(bc)(de)f}
\biggr)
\,.
\equn\label{boxsumEQ}
$$
Here the sum runs over the permutations $P'_6$ and $P''_6$ of indices
$\{123456\}$ modulo symmetries of the integral functions $I_4^{(abc)def}$ and
$I_4^{a(bc)(de)f}$ respectively.
The dimensionally regulated 
integral functions $I_4^{(abc)def}$,~\footnote{See ref.~\cite{BDDKb} for
definitions and conventions.} are symmetric in $a$, $b$ and $c$
and under the exchange of $d$ and $f$. The integral function $I_4^{a(bc)(de)f}$
are symmetric under the exchange of $b$ and $c$ and of $e$ and $f$ independently.

The box coefficients $\hat c$ have been computed in~\cite{BeBbDu} in terms of 
$\NeqFour$ box coefficients $\hat c_{\NeqFour}$.
For example the two mass hard coefficient can be expressed as 
$$
\hat c^{\,a(bc)(de)f}=
2 s_{bc}s_{de}\times\sum_{i=ns,s} \hat c_{\NeqFour,i}^{\,a(bc)(de)f}\hat c_{\NeqFour,i}^{\,a(bc)(ed)f} \,,
\equn\label{KLTboxEQ}
$$
where the label $i$ can take the values ``singlet'' (s) or ``non-singlet'' (ns), as we will discuss presently.

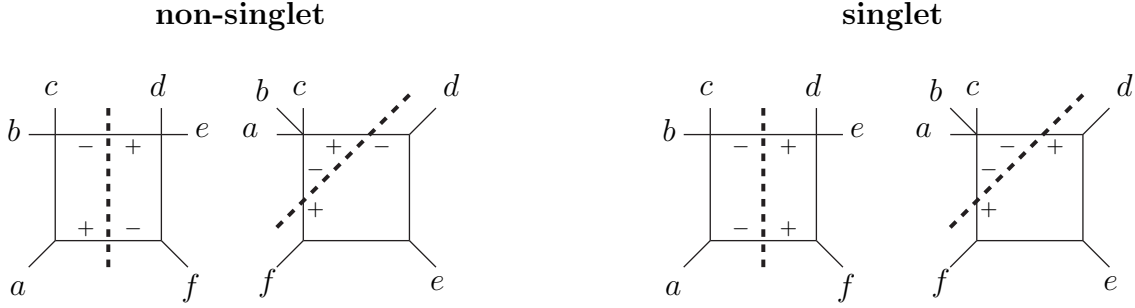
\begin{figure}
\begin{picture}(80,120)(-10,0)

\Line(30,30)(30,70)
\Line(70,30)(70,70)
\Line(30,30)(70,30)
\Line(70,70)(30,70)

\Line(30,30)(20,20)
\Line(70,30)(80,20)

\Line(30,70)(20,70)
\Line(30,70)(30,80)

\Line(70,70)(70,80)
\Line(70,70)(80,70)

\Text(13,12)[l]{$a$}
\Text(78,12)[l]{$f$}
\Text(12,72)[l]{$b$}
\Text(26,88)[l]{$c$}

\Text(66,89)[l]{$d$}
\Text(83,72)[l]{$e$}

\Text(68,115)[l]{\bf non-singlet}

\Text(42,64)[c]{$^-$}
\Text(60,64)[c]{$^+$}
\Text(60,33)[c]{$^-$}
\Text(42,33)[c]{$^+$}

\SetWidth{1.5}
\DashLine(50,20)(50,80){3}
\end{picture}
\begin{picture}(80,120)(-20,0)

\Line(30,30)(30,70)
\Line(70,30)(70,70)
\Line(30,30)(70,30)
\Line(70,70)(30,70)

\Line(30,30)(20,20)
\Line(70,30)(80,20)

\Line(30,70)(20,70)
\Line(30,70)(20,80)
\Line(30,70)(30,80)

\Line(70,70)(80,80)

\Text(13,14)[l]{$f$}
\Text(78,14)[l]{$e$}

\Text(7,72)[l]{$a$}
\Text(12,87)[l]{$b$}
\Text(26,88)[l]{$c$}

\Text(83,89)[l]{$d$}

\Text(35,40)[c]{$^+$}
\Text(35,55)[c]{$^-$}
\Text(42,64)[c]{$^+$}
\Text(60,64)[c]{$^-$}

\SetWidth{1.5}
\DashLine(20,35)(70,85){3}

\end{picture}
\begin{picture}(80,120)(-90,0)

\Line(30,30)(30,70)
\Line(70,30)(70,70)
\Line(30,30)(70,30)
\Line(70,70)(30,70)

\Line(30,30)(20,20)
\Line(70,30)(80,20)

\Line(30,70)(20,70)
\Line(30,70)(30,80)
\Line(70,70)(70,80)
\Line(70,70)(80,70)

\Text(80,115)[l]{\bf singlet}

\Text(13,12)[l]{$a$}
\Text(78,12)[l]{$f$}
\Text(12,72)[l]{$b$}
\Text(26,88)[l]{$c$}
\Text(66,89)[l]{$d$}
\Text(83,72)[l]{$e$}
\Text(42,64)[c]{$^-$}
\Text(60,64)[c]{$^+$}
\Text(60,33)[c]{$^+$}
\Text(42,33)[c]{$^-$}

\SetWidth{1.5}
\DashLine(50,20)(50,80){3}

\end{picture}\hspace{2cm}
\begin{picture}(80,120)(-50,0)

\Line(30,30)(30,70)
\Line(70,30)(70,70)
\Line(30,30)(70,30)
\Line(70,70)(30,70)

\Line(30,30)(20,20)
\Line(70,30)(80,20)

\Line(30,70)(20,70)
\Line(30,70)(20,80)
\Line(30,70)(30,80)
\Line(70,70)(80,80)

\Text(13,14)[l]{$f$}
\Text(78,14)[l]{$e$}

\Text(7,72)[l]{$a$}
\Text(12,87)[l]{$b$}
\Text(26,88)[l]{$c$}

\Text(83,89)[l]{$d$}

\SetWidth{1.5}

\Text(35,40)[c]{$^+$}
\Text(35,55)[c]{$^-$}
\Text(42,64)[c]{$^-$}
\Text(60,64)[c]{$^+$}

\DashLine(20,35)(70,85){3}

\end{picture}

\caption{Examples of non-singlet and singlet contributions to
a two-mass hard and a one-mass box integral. The dashed line indicates the cut to which the
singlet and non-singlet description refer.\label{SingletNonsingletFigure}}
\end{figure}

Six point box coefficients contain two terms which
we label as non-singlet or singlet
$$
\hat c_{} = \hat c_{ns} + \hat c_{s} \,.
\equn\label{SingletPlusNonSinglet}
$$
The two terms arise from different helicity structures in the
cut in three-particle channels, as illustrated in
fig.~\ref{SingletNonsingletFigure}.
The singlet term corresponds to the two cut legs having the same helicity
on one side of the cut and has contributions only
from gluons/gravitons crossing the cut. The non-singlet term has its cut legs
having opposite helicity on one side of the cut.  For this
configuration all terms in the $\NeqFour/\NeqEight$ multiplet contribute.
Note that the designation of singlet/non-singlet depends on which
channel we are considering.

It turns out that two-mass as well as the single mass box coefficients can be expressed in terms
of a single generating function in each case. Specifically,
the two-mass hard coefficients can be generated from
$$
\hat G_0\equiv {i \over 2}
{ s_{bc}s_{de} s_{af}^2 (K_{abc}^2)^8 \over
 \spb{a}.b\spb{b}.c\spb{a}.c\spb{c}.b
\spa{d}.e\spa{e}.f\spa{e}.d\spa{d}.{f}
\BR{a}{d}\BR{a}e
\BR{b}e\BR{c}f    }\,,
\equn
$$
where we are using the usual spinor products $
\spa{j}.{l} \equiv \langle j^- | l^+ \rangle = \bar{u}_-(k_j) u_+(k_l)$
and
$\spb{j}.{l}\equiv  \langle j^+ | l^- \rangle = \bar{u}_+(k_j) u_-(k_l)$, and where
$\BR{i}{j}$ denotes $\BRTTT{i}{j}$ with $K_{abc}^\mu =k_a^\mu+k_b^\mu+k_c^\mu$ and
$s_{ab}=(k_a+k_b)^2$. We are using a spinor helicity basis for the graviton polarisation
tensors~\cite{BerGiKu,GravitySpinor} 
where $\eps^{\pm}_{\mu\nu}\equiv \eps^{\pm}_{\mu}\eps^{\pm}_{\nu}$ and 
$\eps^{\pm}_{\mu}$ are the usual Yang-Mills polarisation vectors~\cite{SpinorHelicity}.

For the coefficients $\hat c^{\,a^-(b^-c^-)(d^+e^+)f^+}$ and
$\hat c^{\,a^+(b^+c^+)(d^-e^-)f^-}$ there is only a singlet contribution and we
have
$$
\eqalign{
\hat c^{\,a^-(b^-c^-)(d^+e^+)f^+}
&=\hat G_0\,,
\cr
\hat c^{\,a^+(b^+c^+)(d^-e^-)f^-}
&=\hat G^*_0\,.
\cr}
\equn
$$
The definition of $\hat G^*_0$ is to parity conjugate $\hat G_0$
by $\spa{a}.b\leftrightarrow \spb{a}.b$
and $\BR{i}{j} \rightarrow \BR{j}{i}$. \footnote{In the 
following we will use the symbols $\hat G_i$ and $\hat H_i$
for the functions $\hat G_i[a,b,c,d,e,f]$ and $\hat H_i[a,b,c,d,e,f]$,
unless the argument is given explicitly.}

The others are sums of a non-singlet and a singlet contribution respectively,
$$
\eqalign{
\hat c^{\,a^+(b^-c^-)(d^+e^+)f^-}
&=\hat G_1 \equiv \left( {\la a | K_{abc} | f \ra
\over K_{abc}^2
}\right)^8
\hat G_0
+
\left( {\spa{b}.c \spb{d}.e
\over K_{abc}^2
}\right)^8
\hat G^*_0\,,
\cr
\hat c^{\,a^-(b^-c^+)(d^-e^+)f^+}
&=
\hat G_2^{} \equiv
\left( { \la c | K_{abc} | d \ra  \over K^2_{abc}  } \right)^8
\hat G_0
+\left( { \spa{a}.b \spb{e}.f \over K^2_{abc}  } \right)^8
\hat G_0^*\,,
\cr
\hat c^{\,a^+(b^-c^+)(d^-e^+)f^-}
&=
\hat G_3^{} \equiv
\left( {  \la e | K_{abc} | b \ra  \over K^2_{abc}  } \right)^8
\hat G_0^*
+\left( { \spb{a}.c \spa{d}.f \over K^2_{abc}  } \right)^8
\hat G_0\,,
\cr
\hat c^{\,a^-(b^-c^+)(d^+e^+)f^-}
&=
\hat G_4^{} \equiv
\left( {  \la c | K_{abc} | f \ra  \over K^2_{abc} } \right)^8
\hat G_0
+\left( { \spa{a}.b \spb{d}.e \over K^2_{abc}  } \right)^8
\hat G_0^*\,,
\cr
\hat c^{\,a^+(b^+c^-)(d^-e^-)f^+}
&=
\hat G_5^{} \equiv
\left( { \la f | K_{abc} | c \ra  \over K^2_{abc}  } \right)^8
\hat G_0^*
+\left( { \spb{a}.b \spa{d}.e \over K^2_{abc}  } \right)^8
\hat G_0\,.
\cr}
\equn
$$
Note that the various terms have the structure implied by (\ref{KLTboxEQ}),
which can be verified by comparing to the expressions of the super-Yang-Mills box
coefficients given in~\cite{BDDKb}.

The one-mass box coefficients can be generated from 

$$
\hat H_0  \equiv {i \over 2}
{\spb{d}.{e}^2\spb{e}.{f}^2  (K^2_{abc})^7 \Bigl(
\spb{a}.b\spa{b}.c \BR{c}{f}   \spa{d}.{a}
+
\spa{a}.{b} \spb{b}.{c}\spa{c}.{d}\BR{a}{f}
\Bigr)
\over
\spb{a}.{b}\spb{b}.{c}\spb{a}.{c}
\BR{a}{d}\BR{a}{f}
\BR{b}d\BR{b}f
\BR{c}d\BR{c}f  }\,\,.
\equn
$$
Again, there are two box coefficients with only a ``singlet'' term,
$$
\eqalign{
\hat c^{\,(a^-b^-c^-)d^+e^+f^+} = \hat H_0\,,
\cr
\hat c^{\,(a^+b^+c^+)d^-e^-f^-} = \hat H^*_0\,.
\cr}
\equn
$$
The remaining box coefficients are sums of non-singlet and singlet terms
$$
\eqalign{
\hat c^{\,(a^-b^+c^+)d^-e^-f^+}&=\hat H_1  \equiv
\left(
{ \BR{f}{a} \over K^2_{abc} }\right)^8
\hat H^*_0
+
\left( { \spb{b}.{c} \spa{d}.e \over K^2_{abc} } \right)^8
\hat H_0\,,
\cr
\hat c^{\,(a^-b^+c^+)d^-e^+f^-}&=\hat H_2  \equiv
\left(
{ \BR{e}{a} \over K^2_{abc} }\right)^8
\hat H^*_0
+
\left( { \spb{b}.{c} \spa{d}.f \over K^2_{abc} } \right)^8
\hat H_0\,,
\cr
\hat c^{\,(a^-b^+c^+)d^+e^-f^-}&=\hat H_3  \equiv
\left(
{ \BR{d}{a} \over K^2_{abc} }\right)^8
\hat H^*_0
+\left( { \spb{b}.{c} \spa{e}.f \over K^2_{abc} } \right)^8
\hat H_0\,,
\cr
\hat c^{\,(a^-b^-c^+)d^-e^+f^+}&=\hat H_4 \equiv
\left(
{ \BR{c}{d} \over K^2_{abc} }\right)^8
\hat H_0+
\left( { \spa{a}.{b} \spb{e}.f \over K^2_{abc} } \right)^8
\hat H^*_0\,,
\cr
\hat c^{\,(a^-b^-c^+)d^+e^-f^+}&=\hat H_5 \equiv
\left(
{ \BR{c}{e} \over K^2_{abc} }\right)^8
\hat H_0
+\left( { \spa{a}.{b} \spb{d}.f \over K^2_{abc} } \right)^8
\hat H^*_0\,,
\cr
\hat c^{\,(a^-b^-c^+)d^+e^+f^-}&=\hat H_6 \equiv
\left(
{ \BR{c}{f} \over K^2_{abc} }\right)^8
\hat H_0
+
\left( { \spa{a}.{b} \spb{d}.e \over K^2_{abc} } \right)^8
\hat H^*_0\,.
\cr}
\equn
$$

One observes that the singlet and the non-singlet terms are proportional to
 the generating functions $\hat H^*_0$ or $\hat H_0$.
As above this pattern of the box-coefficients resembles the pattern for
$\NeqFour$ super-Yang-Mills six-point box coefficients~\cite{BDDKb}.
Furthermore, the relations
$$
\hat c^{\,(abc)def}_{\NeqFour}=\hat c^{\,(def)abc}_{\NeqFour}
=\hat c^{\,a(bc)(de)f}_{\NeqFour}=\hat c^{\,d(ef)(ab)c}_{\NeqFour}
\equn\label{relym}
$$
hold for  $\NeqFour$ box coefficients. As shown in \cite{Bern:1995ix}
three particle factorisation properties imply these relations.
Similarly, in the $\NeqEight$ theory the following identities
$$\eqalign{
\hat c^{\,(abc)def}
+\hat c^{\,(abc)edf} & =
\hat c^{\,a(bc)(de)f}
+\hat c^{\,b(ac)(de)f}
+\hat c^{\,c(ab)(de)f}\,
,
\cr
\hat c^{\,(abc)def}+
\hat c^{\,(abc)dfe}+
\hat c^{\,(abc)edf} &=
\hat c^{\,(def)abc}+
\hat c^{\,(def)acb}+
\hat c^{\,(def)bac}\,, 
\cr}
\equn\label{GravIdentity}
$$
hold. They can be viewed as symmetrisations of the super-Yang-Mills relations.

It is convenient to define $F$-functions which are rescaled scalar 
box integrals~\cite{BDDKb}
$$
I^{a(bc)(de)f} = -{ 2\rg \over s_{af}K^2_{abc} } F^{a(bc)(de)f}\,,
\;\;\;
I^{(abc)def} = -{ 2\rg \over s_{de}s_{ef} } F^{(abc)def}\,.
\equn
$$
We shall use the convention that coefficients of the integrals $I$
are denoted $\hat c$ whilst the coefficients of the $F$ functions are denoted $c$.
Following this convention, we define
$$
 H_i = -{ 2 \over s_{de}s_{ef}} \hat H_i\,\,,\quad\quad
\;\;\;
G_i = -{ 2 \over s_{af}K^2_{abc} } \hat G_i\,.
\equn
$$

\section{IR Singularities}

Gravity amplitudes contain soft divergences.  At one-loop, the expected
form of the soft divergence is~\cite{DunNorB}
$$
\eqalign{
M^{\rm one-loop}_{\eps^{-1}}(1,2,\ldots, n)
=
{i c_\Gamma \kappa^2}
\times
\bigg(
{ \sum_{i<j}  s_{ij} \ln(-s_{ij} )
\over 2\epsilon}
\bigg)
\times M^{\rm tree} (1,2,\ldots, n)\,.
\cr}
\equn\label{IRamplitudeEQ}
$$
This applies for supersymmetric as well as non-supersymmetric theories.

Contributions to these IR singularities can arise from both box and
triangle integral functions. We will argue below, that no triangle functions
contribute to the IR singularities (\ref{IRamplitudeEQ}), since all
singularities are generated from box functions. Furthermore, we will show, that
the coefficient of the $\ln(-K_{abc})/\epsilon$ singularity vanishes among
box functions. These facts give strong support for the conjecture
that triangle functions
are absent in $\NeqEight$ one-loop amplitudes.

The box integrals contain IR singularities
$$
\eqalign{
I^{(abc)def} |_{1/\eps}
&
=
-{2 \over s_{de}s_{ef} } \Bigl[
{ \ln(-s_{de})+\ln(-s_{ef})-\ln(-K^2_{abc}) \over \eps}
\Bigr]\,,
\cr
I^{a(bc)(de)f} |_{1/\eps}
&
=
-{2 \over s_{af} K^2_{abc} } \Bigl[
{ \ln(-s_{af} )+2\ln(-K^2_{abc})-\ln(-s_{bc} )-\ln(-s_{de} ) \over 2\eps}
\Bigr]\,.
\cr}
\equn
$$
When inserted into (\ref{boxsumEQ}) their contributions to the IR of the
six point amplitude, that is (\ref{IRamplitudeEQ}) with $n=6$, can be computed.

For example we find that 26 terms  contribute to the coefficient of $\ln(-s_{12})$
in the six-point amplitude. Explicitly they are
$$\hspace{-0.25cm}
\eqalign{
-&{1 \over 2}\Bigl(
G_5[5,4,3,1,2,6]
+G_5[6,4,3,1,2,5]
+G_5[4,5,3,1,2,6]
+G_5[6,5,3,1,2,4]
\cr
\null & \hskip 9.0 truecm
+G_5[5,6,3,1,2,4]
+G_5[4,6,3,1,2,5] \Bigr)
\cr
&-{1\over 2}\Bigl(
G_0[3,1,2,4,5,6]
+G_0[3,1,2,4,6,5]
+G_0[3,1,2,6,5,4]\Bigr)
\cr
&-{1\over 2}\Bigl(
G_1[4,1,2,5,6,3]
+G_1[5,1,2,4,6,3]
+G_1[6,1,2,5,4,3]
\Bigr)
\cr
\null & \hskip 0.0 truecm
-{1\over 2}\Bigl(
H_0^*[4,5,6,1,2,3]
+H_0^*[4,5,6,2,1,3] \Bigr)
\cr
&+\Bigl(
H_1[3,5,6,1,2,4]
+H_1[3,4,6,1,2,5]
+H_1[3,5,4,1,2,6]
+H_1[3,5,6,2,1,4]
\cr
\null & \hskip 9.0 truecm
+H_1[3,4,6,2,1,5]
+H_1[3,5,4,2,1,6]  \Bigr)
\cr
&+{1\over 2}\Bigl(
G_4[2,3,4,5,6,1]
+G_4[2,3,5,4,6,1]
+G_4[2,3,6,5,4,1]
+G_4[1,3,4,5,6,2]
\cr
\null & \hskip 9.0 truecm
+G_4[1,3,5,4,6,2]
+G_4[1,3,6,5,4,2] \Bigr)\,.
\cr}
\equn\label{BadTreeForm}
$$
Note that the various terms appear with coefficients $\pm1/2$ and $1$.

Although it is not easy to analyse equation (\ref{BadTreeForm}) analytically, 
it can be verified using
computer algebra, that it is of the correct form
$$
s_{12} \times \Mtree\,.
\equn
$$
The expression for $\Mtree$ was determined independently using the KLT-relation~\cite{KLT}
$$
\eqalign{
M_6^{\rm tree}(1,2,3,4,5,6) &=
-is_{12}s_{45}\ A_6^{\rm tree}(1,2,3,4,5,6)(s_{35}A_6^{\rm tree}(2,1,5,3,4,6)
\cr
& \hskip 0cm \null
+(s_{34}+s_{35})\ A_6^{\rm tree}(2,1,5,4,3,6)) + {\cal P}(2,3,4)\,,
\cr}
\equn\label{KLTSix}
$$
where
${\cal P}(2,3,4)$ represents the sum over
permutations of legs $2,3,4$ and $A^\tree_i$ are the tree-level
$i$-point colour-ordered gauge theory partial
amplitudes~\cite{TreeReview}.

We have verified that the box-coefficients yield the correct coefficient of
$\ln(-s_{ab})$ for all choices of $s_{ab}$.

A further check is to test the coefficient
of $\ln(-K_{abc}^2)/\eps$ which should be zero.
Explicitly the coefficient of $\ln(-K_{123}^2)/\eps$ is
$$
\eqalign{
\Bigl(&H_0[1,2,3,4,5,6] +
             H_0[1,2,3,4,6,5] +
             H_0[1,2,3,5,4,6] +
        H_0^*[4,5,6,1,2,3] \cr
&\hspace{9.5cm}+
             H_0^*[4,5,6,1,3,2] +
             H_0^*[4,5,6,2,1,3]\Bigr)
\cr
&        -\Bigl(G_0[1,2,3,4,5,6] +
             G_0[1,2,3,4,6,5] +
             G_0[1,2,3,6,5,4] +
             G_0[2,1,3,4,5,6] 
\cr
&\hspace{-.3cm}
            +G_0[2,1,3,4,6,5] +
             G_0[2,1,3,6,5,4] +
             G_0[3,2,1,4,5,6] +
             G_0[3,2,1,4,6,5] +
             G_0[3,2,1,6,5,4] \Bigr)\,.
\cr}
\equn
$$
It can be shown to vanish by application of the
identities~(\ref{GravIdentity}). By analogous reasoning, 
all $\ln(-K_{abc}^2)/\eps$ terms also cancel in the sum over boxes. We
conclude that the box functions give the full IR singularity structure.

The above expression~(\ref{BadTreeForm})  displays features of the
amplitude, not present in the KLT-form.
For example when considering the twistor structure of NMHV amplitudes, the
tree amplitudes is  expected to have ``coplanar'' support in twistor space.
This can be tested by acting on the expression for $\Mtree$ with the
differential operator $K_{ijkl}$~\cite{Witten:2003nn},
$$
K_{ijkl} =
\spa{i}.{j} \epsilon^{\dot a \dot b}
{ \partial \over \partial \tilde \lambda_k^{\dot a}  }
{ \partial \over \partial \tilde \lambda_l^{\dot b}  }
+{\rm perms}\,. \equn
$$
For gravity amplitudes, one has to act multiple times with the operator
$K_{ijkl}$ in order to annihilate the amplitude. For the six-point amplitude
it was shown in ref.~\cite{BeBbDu} that $K_{ijkl}^3\Mtree=0$.
This annihilation was rather involved to show using the form of the amplitude
generated by the KLT-relations. One reason for this is that individual
terms in the expression for the KLT tree amplitude are not annihilated
individually by $K_{ijkl}^3$ but combine to zero at the final stage.
In the expression for the tree generated from the IR singularities,
each term is individually annihilated by $K_{ijkl}^3$ because the box-coefficients
of the NMHV amplitudes are generically coplanar~\cite{BDKn,Britto:2004tx,BBDP}.
This form for the tree amplitude is thus much closer to a ``CSW''-type expression
for the amplitude.

The CSW formulation~\cite{CSW} of tree level amplitudes in
terms of MHV Parke-Taylor amplitudes~\cite{ParkeTaylor}, interprets
these amplitudes as vertices in twistor space and uses this to construct
amplitudes with any helicity configuration.
Such constructions can be generalised to one-loop MHV calculations~\cite{Brandhuber:2004yw,
Quigley:2004pw,Bedford:2004py}
and for other particle types~\cite{CSW:matter,CSW:massive}. However it still remains
a challenge to give a CSW formulation for gravity tree amplitudes~\cite{GBgravity,ZhuGrav}.
Compact tree level forms, such as the above, are build up from individual terms which
independently are annihilated by $K_{ijkl}^n$, hence such an expression
is a potential 
starting point for formulating
a CSW construction for gravity tree amplitudes.

\section{Connection to Recursion Relations}
The requirement, that the coefficients of the IR singularities
reproduce the tree amplitude, can surprisingly be used to generate
tree amplitudes which are often in a very compact form.  This was
observed in~\cite{BDKn} and used in~\cite{Roiban:2004ix} to obtain a
compact form of one of the eight-point amplitudes with four negative
helicities in $\NeqFour$ Yang-Mills.  By examining the general
structure of these IR relations, Britto, Cachazo and Feng
proposed~\cite{Britto:2004ap} a recursion relation to evaluate tree
amplitudes in Yang-Mills theory.
These recursive relations have been extended to include graviton
scattering~\cite{Bedford:2005yy} where alternate expression for
the MHV graviton amplitudes~\cite{BerGiKu} were found, and
in~\cite{Cachazo:2005ca} where the recursion relations were used to give
an explicit form of the six-point NMHV amplitude.
In this section we shall relate the box-coefficients to the
later computation.

Our starting point is the IR relation for the coefficient of
$\ln(-s_{34})$ within the amplitude
$\Mloop(1^-,2^-,3^-,4^+,5^+,6^+)$,
$$
\sum_{i\in S_{tot}}  a_i c_i   = s_{34} \times \Mtree\,,
\equn
$$
where the sum of $S_{tot}$ is over boxes which have a $\ln(-s_{34})/\epsilon$
singularity. Specifically,
$$
\eqalign{\sum_{i\in S_{tot}}  a_i c_i =&
{1\over 2}\Bigl(
G_0[3,1,2,5,6,4]\Bigr)
+{1\over 2}\Bigl(
G_1[4,1,2,5,6,3]\Bigr)
\cr
&+{1\over 2}\Bigl(
G_3[4,1,5,2,6,3]+
G_3[4,2,5,1,6,3]+
G_3[4,1,6,2,5,3]+
G_3[4,2,6,1,5,3]\Bigr)
\cr
&-{1\over 2}\Bigl(
G_4[2,3,4,5,6,1]+
G_4[1,3,4,5,6,2] \Bigr)
-{1\over 2}\Bigl(
G_5[6,4,3,1,2,5]+
G_5[5,4,3,1,2,6] \Bigr)
\cr
&-{1\over 2}\Bigl(
G_3[6,3,4,2,5,1]+
G_3[6,3,4,1,5,2]+
G_3[5,3,4,2,6,1]+
G_3[5,3,4,1,6,2] \Bigr)
\cr
&-{1\over 2}\Bigl(
G_2[1,3,4,2,5,6]+
G_2[2,3,4,1,5,6]+
G_2[1,3,4,2,6,5]+
G_2[2,3,4,1,6,5] \Bigr)
\cr
&+\Bigl(
H_4[1,2,6,3,4,5]+
H_4[1,2,5,3,4,6] \Bigr)
+\Bigl(
H_2[2,5,6,3,4,1]+
H_2[1,5,6,3,4,2] \Bigr)
\cr
&+\Bigl(
H_5[1,2,6,4,3,5]+
H_5[1,2,5,4,3,6]\Bigr)
+\Bigl(
H_3[2,5,6,4,3,1]+
H_3[1,5,6,4,3,2] \Bigr)\,.
\cr}
\equn
$$
Although this does give a mechanism for generating the tree
amplitude it is not optimally compact.  
Using the identities~(\ref{GravIdentity}), we can reduce this to a smaller expression which will
also generate the tree amplitude, 
$$
\eqalign{\sum_{i\in S}  a_i c_i =&
{1\over 2}\Bigl(
G_0[3,1,2,5,6,4]\Bigr)
+{1\over 2}\Bigl(
G_1[4,1,2,5,6,3]\Bigr)
\cr
&+{1\over 2}\Bigl(
G_3[4,1,5,2,6,3]+
G_3[4,2,5,1,6,3]+
G_3[4,1,6,2,5,3]+
G_3[4,2,6,1,5,3]\Bigr)
\cr
&+{1\over 2}\Bigl(
H_4[1,2,6,3,4,5]+
H_4[1,2,5,3,4,6] \Bigr)
+{1\over 2}\Bigl(
H_2[2,5,6,3,4,1]+
H_2[1,5,6,3,4,2] \Bigr)
\cr
&+{1\over 2}\Bigl(
H_5[1,2,6,4,3,5]+
H_5[1,2,5,4,3,6]\Bigr)
+{1\over 2} \Bigl(
H_3[2,5,6,4,3,1]+
H_3[1,5,6,4,3,2] \Bigr)\,.
\cr}
\equn
$$
analogous to expressions in the Yang-Mills case~\cite{Roiban:2004ix}.

In Yang-Mills theory,
it was noticed that, if we split the box-coefficients into singlet and non-singlet terms, the sum  contains sub-sets which
individually sum to the tree amplitudes. This same simplification also
arises here and we have
$$
\sum_{i\in S'} a_i c_i   =
\sum_{i\in S-S'} a_i c_i   ={1 \over 2} \times s_{34} \times \Mtree
\,,
\equn
$$
with
$$
\eqalign{
\sum_{i\in S'} a_i & c_i   =
{1\over 2}
G_1^{ns}[4,1,2,5,6,3]
\cr
&+{1\over 2}\Bigl(
G_3^{s}[4,1,5,2,6,3]+
G_3^{s}[4,2,5,1,6,3]+
G_3^{s}[4,1,6,2,5,3]+
G_3^{s}[4,2,6,1,5,3]\Bigr)
\cr
&+{1\over 2}\Bigl(
H_4^{ns}[1,2,6,3,4,5]+
H_4^{ns}[1,2,5,3,4,6] \Bigr)
+{1\over 2}\Bigl(
H_2^{s}[2,5,6,3,4,1]+
H_2^{s}[1,5,6,3,4,2] \Bigr)
\cr
&+{1\over 2}\Bigl(
H_5^{s}[1,2,6,4,3,5]+
H_5^{s}[1,2,5,4,3,6]\Bigr)
+{1\over 2} \Bigl(
H_3^{ns}[2,5,6,4,3,1]+
H_3^{ns}[1,5,6,4,3,2] \Bigr)\,.
\cr}
\equn\label{SubSetForm}
$$
This subset corresponds to one quarter of the terms in the full sum.
The terms in this sum are precisely the box-coefficients arising from boxes
with identical helicity structure in the $34$ two-particle cut. Specifically
it corresponds to contributions with intermediate helicity structure 

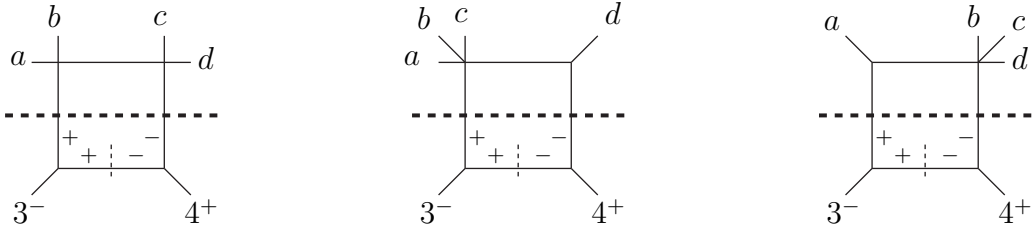
\begin{picture}(150,110)(-20,0)

\Line(30,30)(30,70)
\Line(70,30)(70,70)
\Line(30,30)(70,30)
\Line(70,70)(30,70)

\Line(30,30)(20,20)
\Line(70,30)(80,20)

\Line(30,70)(20,70)
\Line(30,70)(30,80)

\Line(70,70)(70,80)
\Line(70,70)(80,70)

\Text(13,14)[l]{$3^-$}
\Text(78,14)[l]{$4^+$}
\Text(12,72)[l]{$a$}
\Text(26,88)[l]{$b$}

\Text(66,87)[l]{$c$}
\Text(83,72)[l]{$d$}

\Text(35,40)[c]{$^+$}
\Text(66,40)[c]{$^-$}
\Text(60,33)[c]{$^-$}
\Text(42,33)[c]{$^+$}

\DashLine(50,39)(50,27){2}

\SetWidth{1.5}
\DashLine(10,50)(90,50){3}

\end{picture}
\begin{picture}(150,110)(-20,0)

\Line(30,30)(30,70)
\Line(70,30)(70,70)
\Line(30,30)(70,30)
\Line(70,70)(30,70)

\Line(30,30)(20,20)
\Line(70,30)(80,20)

\Line(30,70)(20,70)
\Line(30,70)(20,80)
\Line(30,70)(30,80)

\Line(70,70)(80,80)

\Text(13,14)[l]{$3^-$}
\Text(78,14)[l]{$4^+$}

\Text(7,72)[l]{$a$}
\Text(12,87)[l]{$b$}
\Text(26,88)[l]{$c$}

\Text(83,89)[l]{$d$}

\Text(35,40)[c]{$^+$}

\Text(66,40)[c]{$^-$}
\Text(60,33)[c]{$^-$}
\Text(42,33)[c]{$^+$}

\DashLine(50,39)(50,27){2}

\SetWidth{1.5}
\DashLine(10,50)(90,50){3}

\end{picture}
\begin{picture}(100,110)(-20,0)

\Line(30,30)(30,70)
\Line(70,30)(70,70)
\Line(30,30)(70,30)
\Line(70,70)(30,70)

\Line(30,30)(20,20)
\Line(70,30)(80,20)

\Line(30,70)(20,80)

\Line(70,70)(70,80)
\Line(70,70)(80,80)
\Line(70,70)(80,70)

\Text(13,14)[l]{$3^-$}
\Text(78,14)[l]{$4^+$}
\Text(12,85)[l]{$a$}

\Text(66,88)[l]{$b$}

\Text(83,85)[l]{$c$}
\Text(83,72)[l]{$d$}

\Text(35,40)[c]{$^+$}

\Text(66,40)[c]{$^-$}
\Text(60,33)[c]{$^-$}
\Text(42,33)[c]{$^+$}

\DashLine(50,39)(50,27){2}

\SetWidth{1.5}
\DashLine(10,50)(90,50){3}

\end{picture}

\noindent
This subset of contribution corresponds to contributions where the
trivalent vertex attached to leg three is MHV and the subset of
contributions $S-S'$ is when the trivalent vertex attached to leg four is
MHV.

This summation of terms {\it i.e.,}~eqn.(\ref{SubSetForm}) corresponds
precisely to the terms arising from using the recursive methods of
Britto, Cachazo and Feng~\cite{Britto:2004ap} applied to graviton
scattering ~\cite{Bedford:2005yy,Cachazo:2005ca}.  Specifically in
ref.~\cite{Cachazo:2005ca} the NMHV tree amplitude was written in the
form
$$
D_1+D_1|_{1\leftrightarrow 2}
+\bar D_1
+\bar D_1|_{1 \leftrightarrow 2}
+D_2
+D_2|_{1\leftrightarrow 2}
+D_2|_{5\leftrightarrow 6}
+D_2|_{1,5\leftrightarrow 2,6}
+D_3
+D_3|_{1\leftrightarrow 2}
+\bar D_3
+\bar D_3|_{5\leftrightarrow 6}
+D_6\,,
\equn\label{CachazoSvrcek}
$$ where each term arises from a recursive diagram.  This expression
identifies term-by-term with eq.~(\ref{SubSetForm}) with
$$
\eqalign{
D_1=&H_1^{ns}[1,5,6,2,3,4]/s_{34}\,, \;\;\;\;
D_2=G_3^{s}[4,2,5,1,6,3]/s_{34}\,,
\cr
D_3=&H_2^{s}[2,5,6,1,4,3]/s_{34}\,,\hspace{0.2cm}\;\;\;\;
D_6=G_1^{ns}[4,1,2,5,6,3]/s_{34}\,.
\cr}
\equn
$$
We have checked that this expression
agrees numerically with the expression
for the six-point tree amplitude obtained using the KLT-relation.

By considering, the coefficient of $\ln(-s_{12})$ we can also deduce
that the following subset of terms gives $s_{12}\times \ln(-s_{12})$
$$
\eqalign{
\Bigl(
& H_0^*[4,5,6,1,2,3]
\Bigr)
+\Bigl(
H_1^{ns}[3,5,6,1,2,4]
+H_1^{ns}[3,4,6,1,2,5]
\cr
\null & \hskip -0.3 truecm
+H_1^{ns}[3,5,4,1,2,6]
+H_1^{s}[3,5,6,2,1,4]
+H_1^{s}[3,4,6,2,1,5]
+H_1^{s}[3,5,4,2,1,6]  \Bigr)
\cr
&+\Bigl(
G_4^{ns}[1,3,4,5,6,2]
+G_4^{ns}[1,3,5,4,6,2]
+G_4^{ns}[1,3,6,5,4,2]
\cr
\null & \hskip 4.0 truecm
+G_4^{s}[2,3,4,5,6,1]
+G_4^{s}[2,3,5,4,6,1]
+G_4^{s}[2,3,6,5,4,1]
\Bigr)\,,
\cr}
\equn
$$
which corresponds to the contributions with intermediate helicity structure 

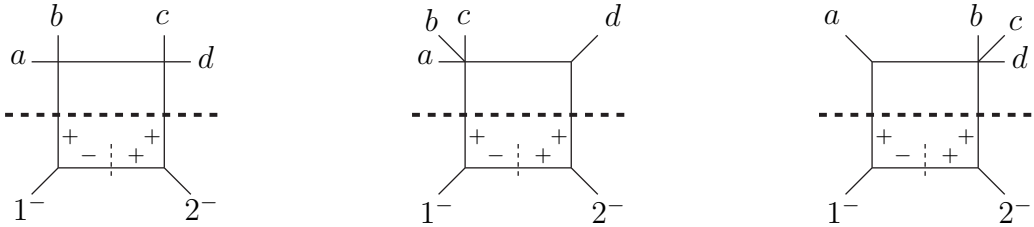
\begin{picture}(150,110)(-20,0)

\Line(30,30)(30,70)
\Line(70,30)(70,70)
\Line(30,30)(70,30)
\Line(70,70)(30,70)

\Line(30,30)(20,20)
\Line(70,30)(80,20)

\Line(30,70)(20,70)
\Line(30,70)(30,80)

\Line(70,70)(70,80)
\Line(70,70)(80,70)

\Text(13,15)[l]{$1^-$}
\Text(78,15)[l]{$2^-$}
\Text(12,72)[l]{$a$}
\Text(27,88)[l]{$b$}

\Text(67,87)[l]{$c$}
\Text(83,72)[l]{$d$}

\Text(35,40)[c]{$^+$}
\Text(66,40)[c]{$^+$}
\Text(60,33)[c]{$^+$}
\Text(42,33)[c]{$^-$}

\DashLine(50,39)(50,27){2}

\SetWidth{1.5}
\DashLine(10,50)(90,50){3}

\end{picture}
\begin{picture}(150,110)(-20,0)

\Line(30,30)(30,70)
\Line(70,30)(70,70)
\Line(30,30)(70,30)
\Line(70,70)(30,70)

\Line(30,30)(20,20)
\Line(70,30)(80,20)

\Line(30,70)(20,70)
\Line(30,70)(20,80)
\Line(30,70)(30,80)

\Line(70,70)(80,80)

\Text(13,14)[l]{$1^-$}
\Text(78,14)[l]{$2^-$}

\Text(12,72)[l]{$a$}
\Text(15,86)[l]{$b$}
\Text(27,87)[l]{$c$}

\Text(83,88)[l]{$d$}

\Text(35,40)[c]{$^+$}
\Text(66,40)[c]{$^+$}
\Text(60,33)[c]{$^+$}
\Text(42,33)[c]{$^-$}

\DashLine(50,39)(50,27){2}

\SetWidth{1.5}
\DashLine(10,50)(90,50){3}

\end{picture}
\begin{picture}(100,110)(-20,0)

\Line(30,30)(30,70)
\Line(70,30)(70,70)
\Line(30,30)(70,30)
\Line(70,70)(30,70)

\Line(30,30)(20,20)
\Line(70,30)(80,20)

\Line(30,70)(20,80)

\Line(70,70)(70,80)
\Line(70,70)(80,80)
\Line(70,70)(80,70)

\Text(13,14)[l]{$1^-$}
\Text(78,14)[l]{$2^-$}
\Text(12,87)[l]{$a$}

\Text(67,88)[l]{$b$}
\Text(82,85)[l]{$c$}
\Text(83,72)[l]{$d$}

\Text(35,40)[c]{$^+$}
\Text(66,40)[c]{$^+$}
\Text(60,33)[c]{$^+$}
\Text(42,33)[c]{$^-$}

\DashLine(50,39)(50,27){2}

\SetWidth{1.5}
\DashLine(10,50)(90,50){3}

\end{picture}

This alternate expression also contains thirteen terms and 
presumably arises by taking legs $1$ and $2$ as the reference
legs for the recursion relations.

\section{Conclusions}

We have presented a form for the box-coefficients of the one-loop NMHV
six-graviton amplitude in $\NeqEight$ supergravity, which makes its
relation to $\NeqFour$ super-Yang-Mills amplitudes manifest.  The
coefficients have a very similar structure to those in $\NeqFour$
super-Yang-Mills being a sum of a singlet and a non-singlet term with
all terms obtained from a generating function.  The IR singularities
contained in these box-functions were shown to give the entire IR
structure of the amplitude. This is strong evidence that the six-point
amplitudes, like those of $\NeqFour$ super-Yang-Mills, are composed
entirely of scalar box-functions with rational coefficients.

We have used the box coefficients to generate expressions for the tree
amplitude. These expressions have a better twistor space structure than
those generated via the KLT relations and could prove to be
related to an underlying CSW type formulation of gravity scattering
amplitudes.

Given the absence of integral functions beyond scalar box integrals
within the five and six-point amplitudes, it becomes difficult to see
how these functions can appear in higher point amplitudes whilst still
satisfying factorisation and soft limit constraints.  Hence, it seems
increasingly likely that $\NeqEight$ supergravity one-loop amplitudes
are composed only of box integral functions.

This simplification is unexpected from power counting arguments, which
are based, on the known symmetries of $\NeqEight$ supergravity.
One might suspect, this implies
the existence of further symmetries and additional
constraints on the scattering amplitudes.  It seems promising,
although a challenge, to utilise the simplification of the one-loop
amplitudes to determine the ultra-violet behaviour of higher loop
scattering amplitudes in $\NeqEight$ supergravity.

We would like to thank Zvi Bern, James Bedford, Andi Brandhuber, Warren Perkins, 
Bill Spence and 
Gabriele Travaglini for useful discussions.
This research was supported by the PPARC and the EPSRC.


\begin{thebibliography}{99}
\bibitem{ExtendedSugra}
E.~Cremmer, B.~Julia and J.~Scherk,
Phys.\ Lett.\ B {\bf 76}, 409 (1978);\\
%
E.~Cremmer and B.~Julia,
Phys.\ Lett.\ B {\bf 80}, 48 (1978).

\bibitem{BDDKa}
Z. Bern, L.J. Dixon, D.C. Dunbar and D.A. Kosower,
\npb{425}{1994}{217}, \hepph{9403226}.
\bibitem{BDDKb}
Z. Bern, L.J. Dixon, D.C. Dunbar and D.A. Kosower,
\npb{435}{1995}{59}, \hepph{9409265}.



\bibitem{BeDeDiKo}
Z.~Bern, V.~Del Duca, L.J.~Dixon and D.~A.~Kosower,
 Phys.\ Rev.\ D {\bf 71}, 045006 (2005)
[hep-th/0410224].




\bibitem{BDKn}
Z.~Bern, L.~J.~Dixon and D.~A.~Kosower,
hep-th/0412210.


\bibitem{Witten:2003nn}
E.~Witten,
Commun.\ Math.\ Phys.\  {\bf 252}, 189 (2004)
[hep-th/0312171].


\bibitem{Cachazo:2004dr}
F.~Cachazo,
hep-th/0410077.


\bibitem{Britto:2004nj}
R.~Britto, F.~Cachazo and B.~Feng,
hep-th/0410179.

\bibitem{Bidder:2004tx}
  S.~J.~Bidder, N.~E.~J.~Bjerrum-Bohr, L.~J.~Dixon and D.~C.~Dunbar,
  Phys.\ Lett.\ B {\bf 606}, 189 (2005)
[hep-th/0410296].

\bibitem{BrittoUnitarity}
R.~Britto, F.~Cachazo and B.~Feng,
hep-th/0412103.

\bibitem{KLT}
H.~Kawai, D.~C.~Lewellen and S.~H.~H.~Tye,
Nucl.\ Phys.\ B {\bf 269}, 1 (1986).


\bibitem{Donoghue:1994dn}
S.~Weinberg,
Physica A{\bf  96}, 327 (1979);\\
J.~F.~Donoghue,
Phys.\ Rev.\ D {\bf 50} 3874 (1994)
[gr-qc/9405057];\\
N.~E.~J.~Bjerrum-Bohr, J.~F.~Donoghue and B.~R.~Holstein,
Phys.\ Rev.\ D {\bf 67}, 084033 (2003)
[hep-th/0211072];
Phys.\ Rev.\ D {\bf 68}, 084005 (2003)
[hep-th/0211071].


\bibitem{EffKLT}
N.~E.~J.~Bjerrum-Bohr,
Phys.\ Lett.\ B {\bf 560}, 98 (2003) [hep-th/0302131];
Nucl.\ Phys.\ B {\bf 673}, 41 (2003) [hep-th/0305062];\\
N.~E.~J.~Bjerrum-Bohr and K.~Risager,
Phys.\ Rev.\ D {\bf 70}, 086011 (2004) [hep-th/0407085].


\bibitem{BDWGravity}
Z.~Bern, A.~De Freitas and H.~L.~Wong,
Phys.\ Rev.\ Lett.\  {\bf 84}, 3531 (2000)
[hep-th/9912033].


\bibitem{GSB}
{M.B.\ Green, J.H.\ Schwarz and L.\ Brink,
 Nucl.\ Phys.\ B198:472 (1982)}.


\bibitem{PassVelt}
G. Passarino and M. Veltman, Nucl. Phys. B {\bf 160}, 151, (1979).


\bibitem{GravityStringBased}
Z. Bern, D.C. Dunbar and T. Shimada, 
  Phys.\ Lett.\ B {\bf 312}, 277, (1993)
[hep-th/9307001];\\
D.C. Dunbar and P.S. Norridge,
\npb{433}{1995}{181} [hep-th/9408014].


\bibitem{Bern:1998sv}
Z.~Bern, L.~J.~Dixon, M.~Perelstein and J.~S.~Rozowsky,
Nucl.\ Phys.\ B {\bf 546}, 423 (1999)
[hep-th/9811140].



\bibitem{Grisaru:1979re}
M.~T.~Grisaru and J.~Zak,
Phys.\ Lett.\ B {\bf 90}, 237 (1980).


\bibitem{Dunbar:1999nj}
 D.~C.~Dunbar, B.~Julia, D.~Seminara and M.~Trigiante,
 JHEP {\bf 0001}, 046 (2000)
 [hep-th/9911158];\\
 D.~C.~Dunbar and N.~W.~P.~Turner,
 Class.\ Quant.\ Grav.\  {\bf 20}, 2293 (2003)
 [hep-th/0212160].


\bibitem{BDDPR}
Z. Bern, L.J. Dixon, D.C. Dunbar, M.\ Perelstein and J.S.\ Rozowsky,
\npb{530}{1998}{401} \hepth{9802162};
\cqg{17}{2000}{979}
\hepth{9911194}.

\bibitem{Howe:2002ui}
  P.~S.~Howe and K.~S.~Stelle,
  Phys.\ Lett.\ B {\bf 554}, 190 (2003)
  [hep-th/0211279].



\bibitem{BeBbDu}
Z.~Bern, N.~E.~J.~Bjerrum-Bohr and D.~C.~Dunbar,
hep-th/0501137.


\bibitem{Bedford:2005yy}
J.~Bedford, A.~Brandhuber, B.~Spence and G.~Travaglini,
  hep-th/0502146.


\bibitem{Cachazo:2005ca}
 F.~Cachazo and P.~Svrcek,
  hep-th/0502160.



\bibitem{BerGiKu}
F.~A.~Berends, W.~T.~Giele and H.~Kuijf,
Phys.\ Lett.\ B {\bf 211}, 91 (1988).




\bibitem{GravitySpinor}
D.~Spehler and S.~F.~Novaes,
  Phys.\ Rev.\ D {\bf 44}, 3990 (1991);
  Nucl.\ Phys.\ B {\bf 371}, 618 (1992);\\
  H.~T.~Cho and K.~L.~Ng,
  Phys.\ Rev.\ D {\bf 47}, 1692 (1993).
                                                                                



\bibitem{SpinorHelicity}
Z.~Xu, D.~H.~Zhang and L.~Chang,
Nucl.\ Phys.\ B {\bf 291}, 392 (1987).



\bibitem{Bern:1995ix}
Z.~Bern and G.~Chalmers,
Nucl.\ Phys.\ B {\bf 447}, 465 (1995)
[hep-ph/9503236].



\bibitem{DunNorB}
D.C. Dunbar and P.S. Norridge, Class. Quantum Grav. {\bf 14}, 351 {(1997)},
[hep-th/9512084].


\bibitem{TreeReview}
M.~L.~Mangano and S.~J.~Parke,
Phys.\ Rept.\  {\bf 200}, 301 (1991);\\
%
L.~J.~Dixon,
in {\it QCD \& Beyond: Proceedings of TASI '95},
ed. D.\ E.\ Soper (World Scientific, 1996) [hep-ph/9601359].




\bibitem{Britto:2004tx}
R.~Britto, F.~Cachazo and B.~Feng,
hep-th/0411107.




\bibitem{BBDP}
S.~J.~Bidder, N.~E.~J.~Bjerrum-Bohr, D.~C.~Dunbar and
W.~B.~Perkins,
hep-th/0412023;
hep-th/0502028.




\bibitem{CSW}
F.~Cachazo, P.~Svrcek and E.~Witten,
JHEP {\bf 0409}, 006 (2004) [hep-th/0403047].


\bibitem{ParkeTaylor}
S.J. Parke and T.R. Taylor,
Phys.\ Rev.\ Lett.\ 56:2459
(1986);\\
%
F.~A.~Berends and W.~T.~Giele,
Nucl.\ Phys.\ B {\bf 306}, 759 (1988).


\bibitem{Brandhuber:2004yw}
A.~Brandhuber, B.~Spence and G.~Travaglini,
 Nucl.\ Phys.\ B {\bf 706}, 150 (2005) [hep-th/0407214].



\bibitem{Quigley:2004pw}
C.~Quigley and M.~Rozali,
 JHEP {\bf 0501}, 053 (2005) [hep-th/0410278].


\bibitem{Bedford:2004py}
J.~Bedford, A.~Brandhuber, B.~Spence and G.~Travaglini,
 Nucl.\ Phys.\ B {\bf 706}, 100 (2005)
 [hep-th/0410280].


\bibitem{CSW:matter}
G.~Georgiou and V.~V.~Khoze,
JHEP {\bf 0405}, 070 (2004)
[hep-th/0404072];\\
%
%
J.~B.~Wu and C.~J.~Zhu,
JHEP {\bf 0409}, 063 (2004) [hep-th/0406146];\\
%
%
X.~Su and J.~B.~Wu,
hep-th/0409228.
%


\bibitem{CSW:massive}
L.~J.~Dixon, E.~W.~N.~Glover and V.~V.~Khoze,
JHEP {\bf 0412}, 015 (2004)
[hep-th/0411092];\\
Z.~Bern, D.~Forde, D.~A.~Kosower and P.~Mastrolia,
hep-ph/0412167;\\
S.~D.~Badger, E.~W.~N.~Glover and V.~V.~Khoze,
hep-th/0412275.




\bibitem{GBgravity}
S.~Giombi, R.~Ricci, D.~Robles-Llana and D.~Trancanelli,
JHEP {\bf 0407}, 059 (2004)
[hep-th/0405086].

\bibitem{ZhuGrav}
J.~B.~Wu and C.~J.~Zhu,
JHEP {\bf 0407}, 032 (2004)
[hep-th/0406085].



\bibitem{Roiban:2004ix}
R.~Roiban, M.~Spradlin and A.~Volovich,
hep-th/0412265.


\bibitem{Britto:2004ap}
R.~Britto, F.~Cachazo and B.~Feng,
hep-th/0412308; \\
R.~Britto, F.~Cachazo, B.~Feng and E.~Witten,
hep-th/0501052.









\end{thebibliography}
\end{document}